\documentclass[letter]{aa}
\usepackage{natbib}
\usepackage{graphicx}
\usepackage{txfonts}
\usepackage{color}
\usepackage{booktabs}
\usepackage[breaklinks,colorlinks=true,citecolor=blue,linkcolor=magenta,urlcolor=blue]{hyperref}
\usepackage{caption}
\usepackage{subfigure}%
\usepackage{rotating}%
\DeclareCaptionLabelFormat{continued}{#1~#2 (continued)}%
\newcommand{\msun}{$M_{\odot}$}
\newcommand{\rsun}{$R_{\odot}$}

\newcommand{\teff}{$T_\mathrm{eff}$}
\newcommand{\logg}{$\log g$}
\newcommand{\logy}{$\log n(\mathrm{He})/n(\mathrm{H})$}

\newcommand{\sdo}{\object{Gaia\,DR2\,5694207034772278400}}
\newcommand{\sd}{J0809-2627}

\newcommand{\kms}{km\,s$^{-1}$}
\newcommand{\vsini}{$v_\mathrm{rot}\sin i$}

\begin{document}

%

\title{Discovery of a highly magnetic He-sdO star from a double-degenerate binary merger}

\author{M.\ Dorsch\inst{1}, N.\ Reindl\inst{2},  I.\ Pelisoli\inst{3},  U.\ Heber\inst{1}, S.\ Geier\inst{2}, A.\ G.\ Istrate\inst{4},  S.\ Justham\inst{5,6,7}}

\institute{
Dr.\ Karl Remeis-Observatory \& ECAP, Friedrich-Alexander University Erlangen-N\"{u}rnberg,
Sternwartstr.\ 7, 96049 Bamberg, Germany\\
\email{matti.dorsch@fau.de} 
\and
Institut für Physik und Astronomie, Universität Potsdam, Haus 28, Karl-Liebknecht-Str. 24/25, 14476 Potsdam-Golm, Germany
\and
Department of Physics, University of Warwick, Gibbet Hill, Road, Coventry, CV4 7AL, UK
\and
Department of Astrophysics/IMAPP, Radboud University, P.\ O.\ Box 9010, NL-6500 GL Nijmegen, The Netherlands
\and
University, and National Observatory, of the Chinese Academy of Sciences, Beijing 100012, China
\and
Anton Pannekoek Institute for Astronomy \& GRAPPA, University of Amsterdam, Postbus 94249, 1090 GE Amsterdam, The Netherlands
\and
Max-Planck-Institut für Astrophysik, Karl-Schwarzschild-Straße 1, 85741 Garching, Germany
}

\date{Received ; accepted }  

\abstract
{
Helium-rich hot subdwarf stars of spectral type O (He-sdO) are considered prime candidates for stellar merger remnants. Such events should lead to the generation of strong magnetic fields.  
However, no magnetic He-sdO has yet been unambiguously discovered despite the high magnetic rate (20\,\%) among white dwarf stars, the progeny of hot subdwarfs. 
Here we present the discovery of a strong magnetic field ($B=353\pm10$\,kG) from Zeeman-split hydrogen, helium, and metal lines in the optical X-SHOOTER spectrum of an He-sdO and present the first spectroscopic analysis of any magnetic hot subdwarf.
For this we used line-blanketed \textsc{Tlusty} non-local thermodynamic equilibrium models and assumed a simple homogeneous magnetic field. The derived atmospheric parameters \teff\ = $44900\pm1000$\,K and \logg\ = $5.93\pm 0.15$ are typical for He-sdO stars, while the star is less hydrogen-poor than most He-sdOs at \logy\ = $+0.28\pm 0.10$. The star is a slow rotator (\vsini\ $<$ 40\,\kms).\ Its chemical composition is N-rich and C- and O-poor, and  the Si and S abundances are close to solar.
Combining the atmospheric parameters with \textit{Gaia} parallax and photometry,
the stellar radius and luminosity are found to be typical for He-sdOs and place the star on the helium main sequence in the Hertzsprung-Russell diagram. Its mass of $0.93^{+0.44}_{-0.30}$\,$M_\odot$, although uncertain, appears to be remarkably high.
The strong magnetic field along with the atmospheric parameters and metal abundances provide overwhelming evidence for the double-degenerate merger scenario. 
}


\authorrunning{Dorsch et al.}
\titlerunning{Discovery of a highly magnetic He-sdO}

\maketitle

\section{Introduction}

Most hot subdwarf-O (sdO) and subdwarf-B (sdB) stars are helium-burning stars with hydrogen envelopes that are too thin to sustain hydrogen shell burning \citep{heber09,heber16}. The bulk of sdB stars form the hot extreme of the horizontal branch at masses close to half solar and radii of 0.1\,--\,0.2 \rsun. %
Various types of binary interaction can explain the majority of hot subdwarf stars \citep{Pelisoli2020}.
This includes stable Roche overflow to a main sequence companion star, common-envelope ejection by white dwarf (WD) companions to red giant stars, and the merging of two helium-core white dwarfs \citep[He-WDs;][]{han02,Han2003,webbink84}. 
Because hot subdwarf stars have very thin envelopes, they evolve directly to the WD sequence without an excursion to the asymptotic giant branch. A smoking gun for the merger scenario would be the detection of strong magnetic fields, which are predicted by detailed merger models for massive stars \citep{schneider2019} and carbon-oxygen (CO) WD mergers \citep{Ji2013,Zhu2015} and are likely to occur for helium WD mergers as well.

About 20\,\% of all WD stars within 20\,pc of Earth are known to host surface magnetic fields, with strengths ranging from a few kilogauss to several hundred megagauss \citep[e.g.][]{Bagnulo2020,Bagnulo2021}.
In contrast, magnetic fields have not yet been directly observed for any hot subdwarf star.
Spectropolarimetry has failed to provide evidence for the magnetic fields of more than 40 sdB and sdO stars, with upper limits of about 1 to 2\,kG \citep{Landstreet2012,Mathys2012,Randall2015,Bagnulo2015}.
Indirect signs of magnetism have been claimed from light and radial velocity variations tentatively attributed to magnetic spots \citep{Jeffery2013,Geier2015,Balona2019,Momany2020,Vos2021a}, but lack independent confirmation.

A first helium-rich hot O-type subdwarf (He-sdO) star with a significant magnetic field ($\approx$300\,--\,700\,kG) was identified spectroscopically by \cite{Heber2013}, although this star has not been analysed to date.  
Here we present the first spectroscopic analysis of a hot subdwarf star with a strong magnetic field, the He-sdO \sdo\  (henceforth \sd).  

\section{Observations}

\sd\ was identified as a candidate hot subdwarf by \cite{Geier2019}. We obtained follow-up spectroscopy with the Intermediate Dispersion Spectrograph (IDS) spectrograph at the Isaac Newton Telescope (INT) to confirm this classification and identified the star as a helium-rich sdO star. 
At a resolution of $\Delta\lambda$ $\approx$ 4\,\AA\ and signal-to-noise ratio (S/N) of 50, the IDS spectrum %
shows strongly broadened hydrogen and helium lines, %
which we tentatively interpreted as rapid rotation.

\begin{figure}
\centering
\includegraphics[width=0.49\textwidth]{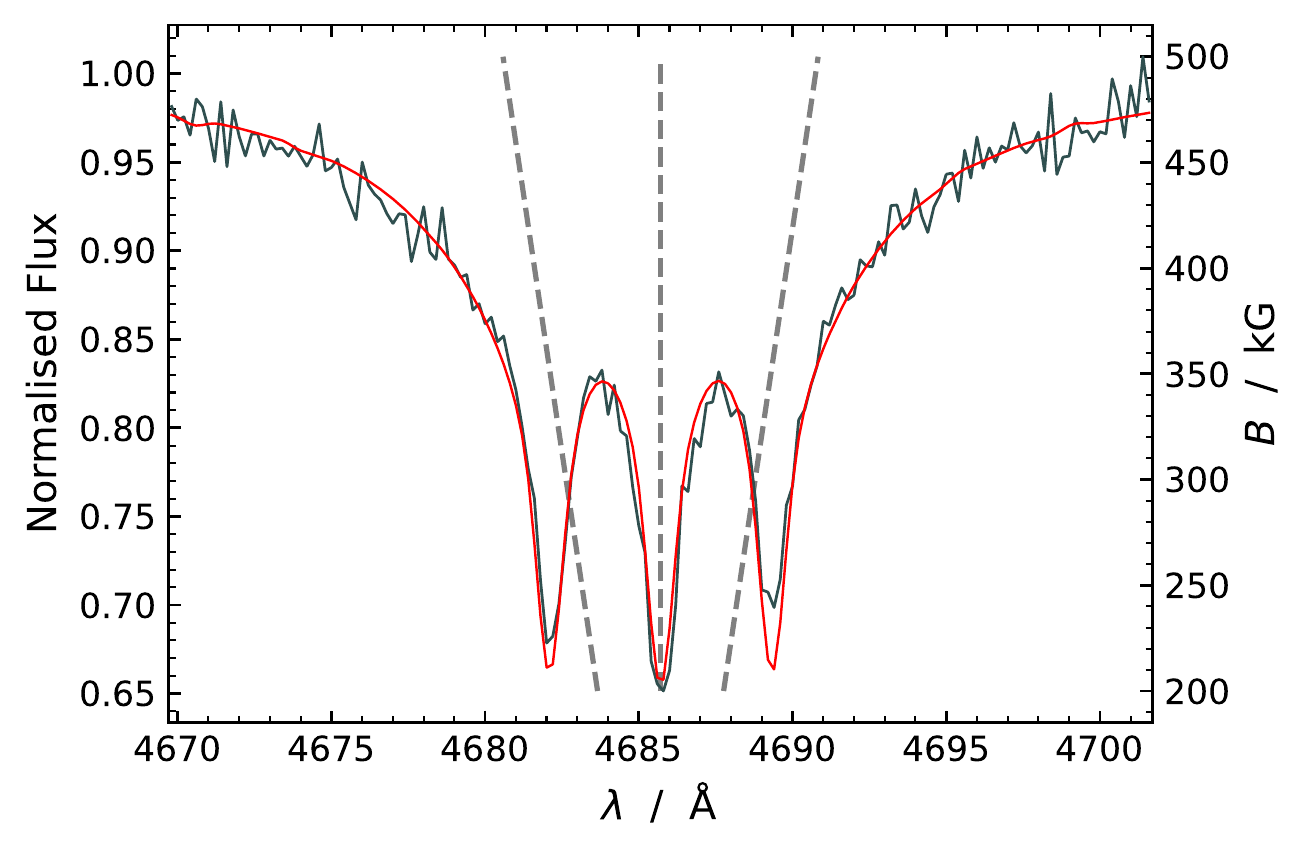}
\caption{Zeeman-split \ion{He}{ii}\,4686\,\AA\ line in the radial-velocity-corrected X-SHOOTER spectrum of \sd\ (grey). The best-fit model is shown in red. The dashed lines illustrate the positions of the three components that depend on the magnetic field strength due to the linear Zeeman effect.}
\label{fig:he2}
\end{figure}

Because hot subdwarf stars are known to be slow rotators \citep{geier2012}, the strong line-broadening of \sd\ is unusual.
Therefore, the star was observed again in April 2021 with the medium-resolution spectrograph X-SHOOTER at the ESO-VLT.
We used the 3070\,--\,7400\,\AA\ range of the UVB and VIS channels, which covers all relevant spectral lines at a mean S/N of about 80 and at a resolving power of  $R$\,=\,$\lambda / \Delta \lambda$\,$\approx$\,$10000$. 
At this resolution, which is eight times better than that of IDS, strong Zeeman-split multiplets are clearly visible, demonstrating that the line broadening is not due to rapid rotation but is caused by the presence of a magnetic field (see Fig.~\ref{fig:he2}).
The 3070 -- 7400\,\AA\ spectral range is shown in Fig.~\ref{appendix:xshooter}. 
All exposure times can be found in Appendix \ref{sect:appendix_obs_spec}.

\begin{figure*}
\centering
\includegraphics[width=0.99\textwidth]{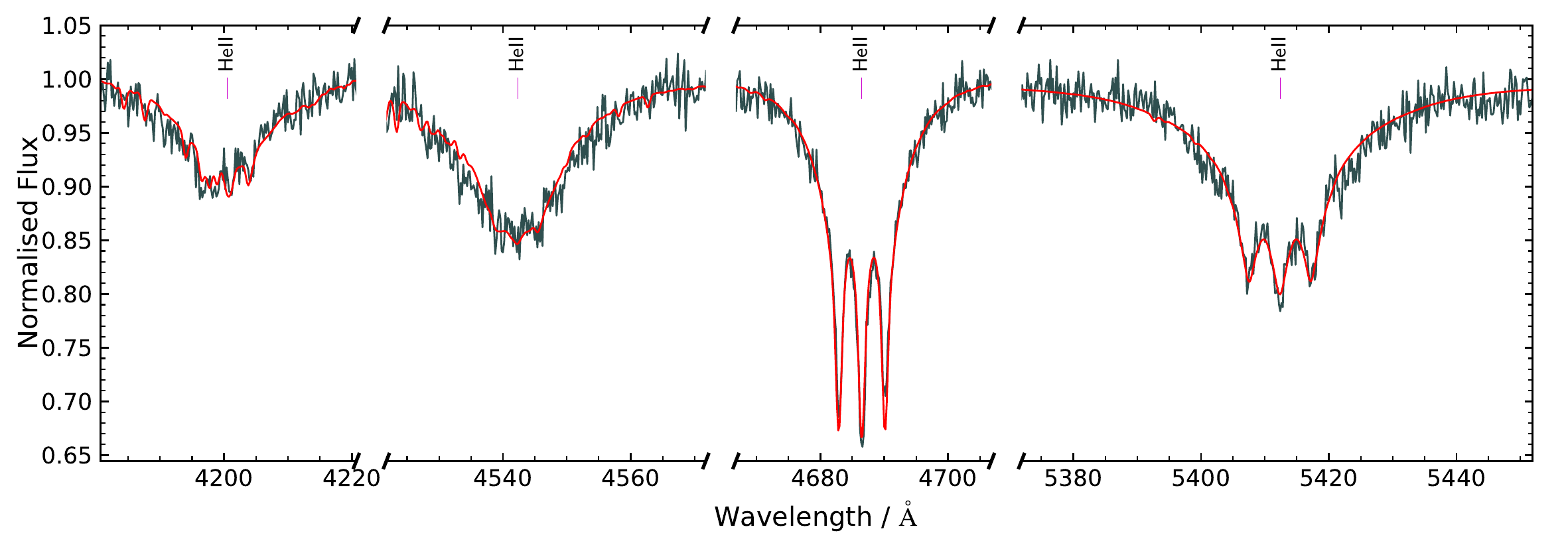}\vspace{-20pt}
\includegraphics[width=0.99\textwidth]{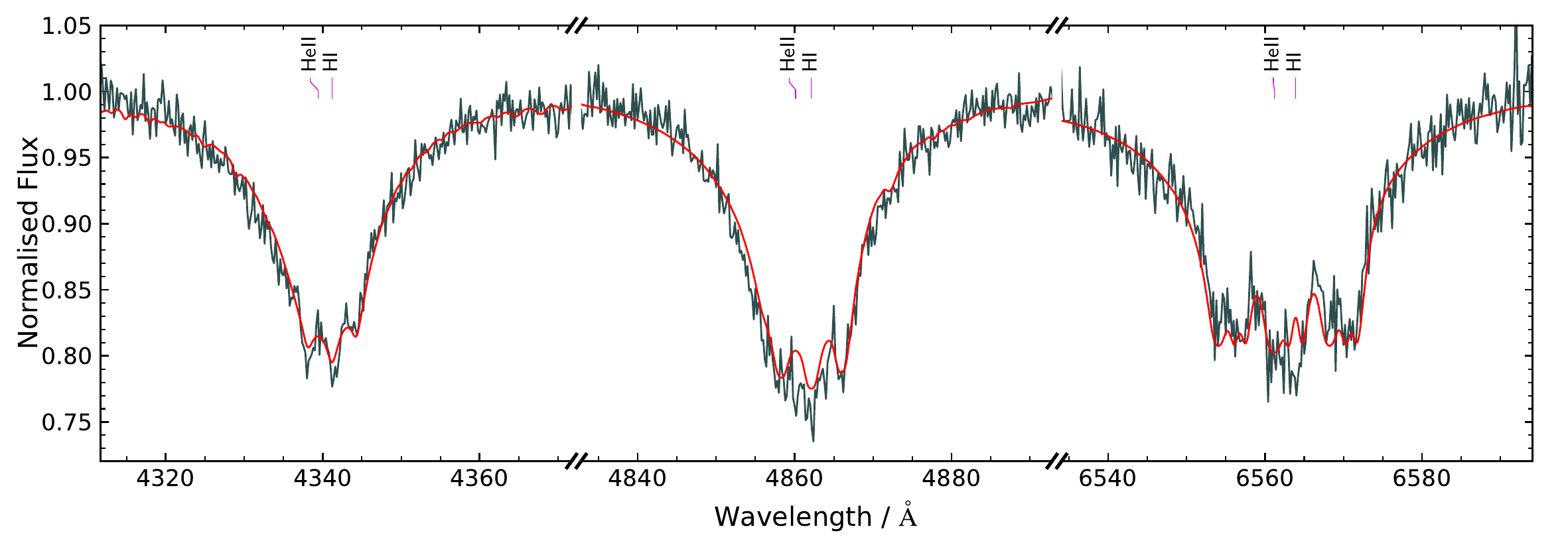}\vspace{-20pt}
\includegraphics[width=0.99\textwidth]{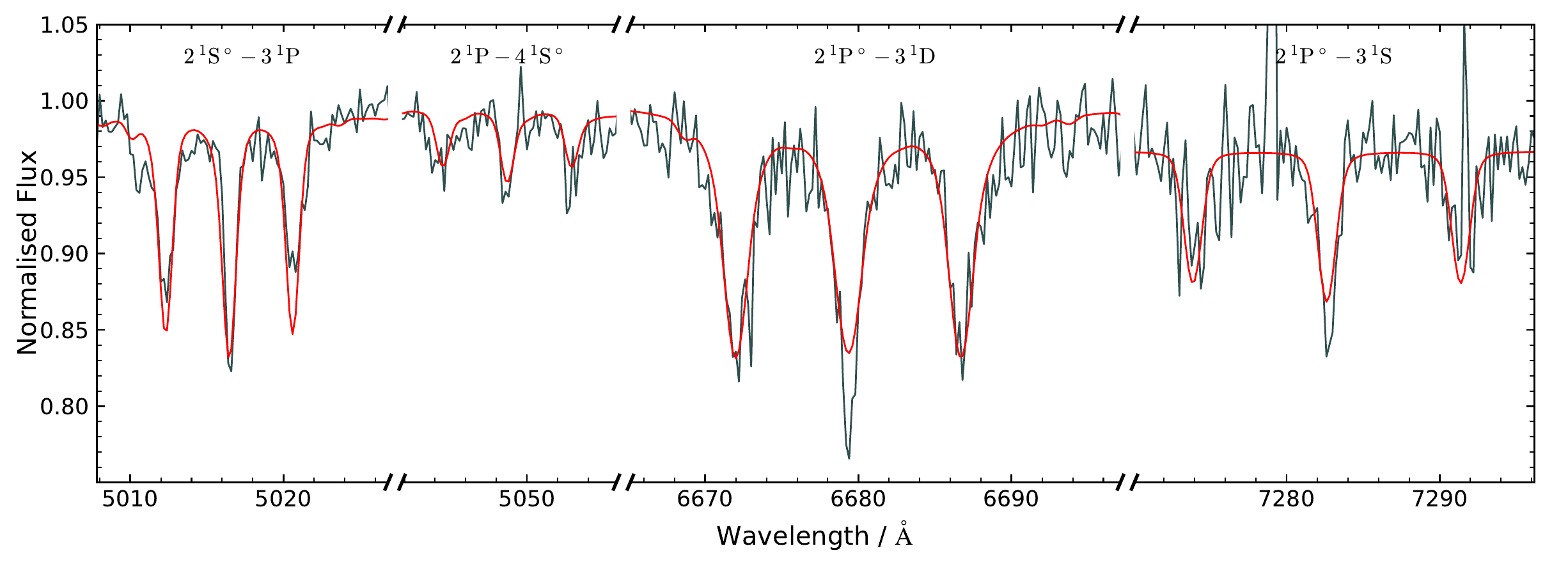}\vspace{-20pt}
\includegraphics[width=0.99\textwidth]{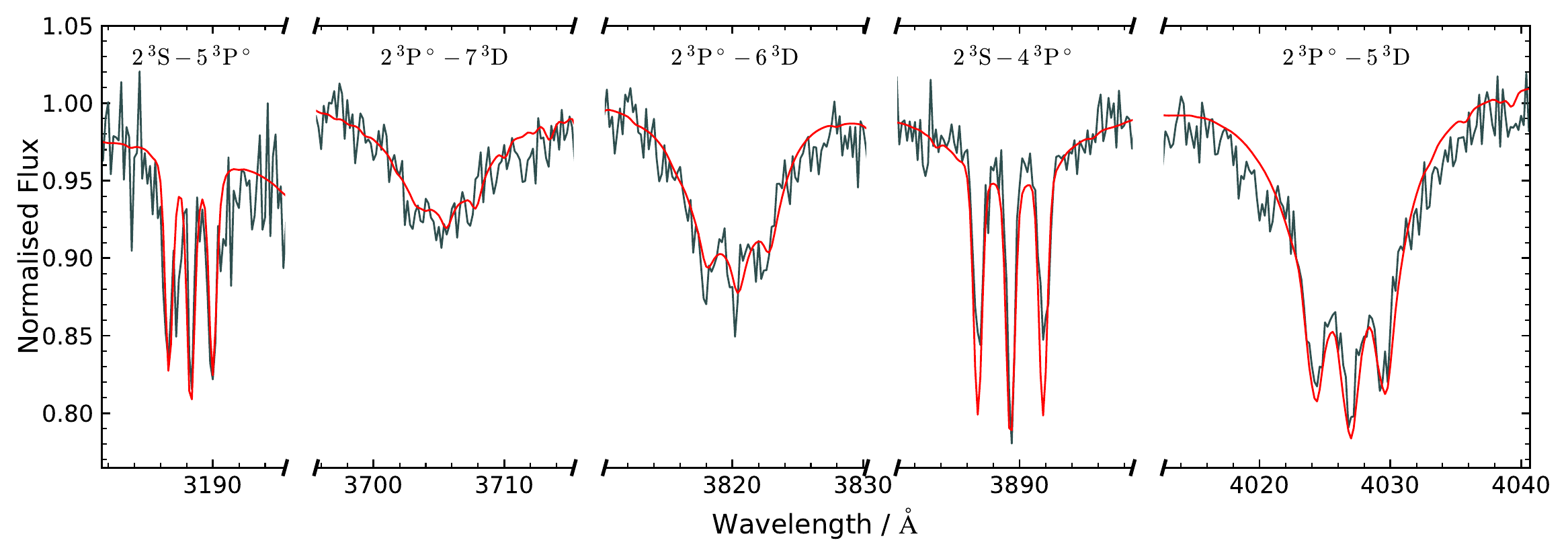}
\caption{
Strongest \ion{He}{ii} and \ion{H}{i} lines, as well as the best-fitting \ion{He}{i} lines in the X-SHOOTER spectrum of \sd\ (grey).
The best-fit model is shown in red.
Labels indicate \ion{He}{ii} and \ion{H}{i} line positions at $B=0$, as well as lower and upper LS terms for \ion{He}{i}.
}
\label{fig:atmlines}
\end{figure*}

The light curve of \sd\ has been observed by the \textit{TESS} satellite but does not show significant variability at a $\approx$~0.42~parts per thousand (ppt) detection limit (see Appendix \ref{sect:appendix_obs_tess}). However, \sd\ is located in a relatively crowded field, so contamination by nearby stars may hide any variability. 

\section{Models}

We used a grid of atmospheric structures computed with the \textsc{Tlusty} code as the basis of our spectroscopic analysis. A description of the code is given in \cite{hubeny17a,hubeny17b,hubeny17c}. 
These models are plane-parallel, homogeneous, hydrostatic, and include H, He, C, N, O, Si, P, S, Fe, and Ni in non-local thermodynamic equilibrium.

The observed line splittings can be explained by the linear Zeeman effect caused by the presence of a strong surface magnetic field.
The magnetic field was assumed to be homogeneous across the visible surface, that is, of uniform strength and direction. Its axis was allowed to be inclined at an angle $\psi$ with respect to the line of sight.
Smaller inclinations lead to stronger displaced Zeeman components relative to the central component.
The magnetic field in \sd\ is not strong enough to have a large effect on the atmospheric structure beyond the additional metal line opacity in the UV region \citep{Wickramasinghe1986, Tremblay2015}\footnote{
Although iron and nickel lines are not observed in the optical spectrum of \sd, they were still included in our models at abundances of three times their solar value. This is necessary because the opacity of Zeeman-spilt iron group lines in the UV region likely has a significant effect on the atmospheric structure. 
For the same reason, we used a microturbulent velocity of 6\,\kms\ for the atmospheric structure calculations only.}. 
In addition, the linear Zeeman effect dominates over the quadratic Zeeman effect at strengths below about 1\,MG \citep{Garstang1974}. 
Therefore, our approach was to consider  only the linear Zeeman effect and only in the spectrum synthesis.
We modified the spectrum synthesis code \textsc{Synspec} \citep{hubeny17a} to include linear Zeeman multiplets for hydrogen, helium, and detectable metal lines, similar to the method \cite{Kawka2011} used to model the cool magnetic WD NLTT\,10480, which has a magnetic field strength ($\approx$500\,kG) similar to that of \sd. 
Polarised radiative transfer in the lines was not included in our simple models.
More details of our model for the magnetic field are given in Appendix \ref{appendix:zeeman}.

\section{Spectral fits}

\begin{table}
\caption{Atmospheric parameters derived from spectroscopy. Elemental abundance results are stated by number fraction. 
Upper limits are given as best-fit values, while their uncertainties represent values that can clearly be excluded.
}
\vspace{-10pt}
\label{tab:results}
\setstretch{1.2}
\begin{center}
\begin{minipage}{0.3\textwidth}
\begin{tabular}{l @{} r}
\toprule
\toprule
{} & Spectral fit   \\
\midrule
$T_\mathrm{eff}$ (K)    & $44900 \pm 1000$ \\
$\log g$   & $\phantom{+}5.93 \pm 0.15$  \\
$\log n(\mathrm{He}) / n(\mathrm{H})$  & $+0.28 \pm 0.10$  \\
$B$ (kG)  & $353 \pm 10$  \\
$\psi$ ($\degr$)  & $\phantom{0}64\pm25$  \\
$v_\mathrm{rad}$ (km\,s$^{-1}$) & $\phantom{0}33\pm2$  \\
$v_\mathrm{rot}\sin i~$ (km\,s$^{-1}$) & $<40$  \\
\bottomrule
\end{tabular}
\end{minipage}
\hspace{-20pt}
\begin{minipage}{0.2\textwidth}
\begin{tabular}{l @{} r} %
\toprule
\toprule
 Element  & \multicolumn{1}{c}{$\log n_\mathrm{X} / \sum_i n_i$} \\ %
\midrule
H       &                                     $-0.46\pm0.07$ \\%
He      &                                     $-0.18\pm0.03$ \\%
C       &                         <$-3.74^{+0.40}_{}$ \\%
N       &                                     $-2.98\pm0.20$ \\%
O       &                         <$-3.92^{+0.30}_{}$ \\%
Si      &                                      $-4.26\pm 0.30$ \\%
S       &                                     $-4.16\pm0.30$ \\%
\bottomrule
\end{tabular}
\end{minipage}
\end{center}
\end{table}

\begin{figure*}%
\centering
\includegraphics[width=0.99\textwidth]{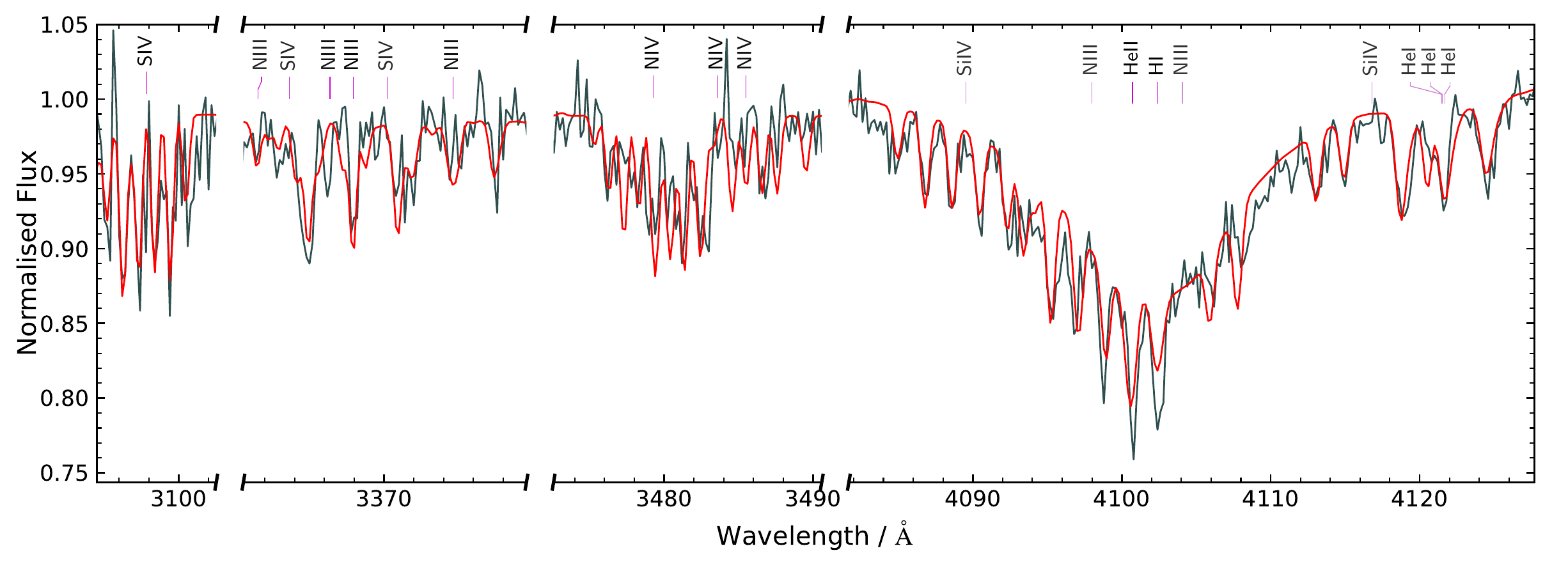}
\caption{Zeeman-split \ion{N}{iii-iv}, \ion{Si}{iv}, and \ion{S}{iv} lines in the X-SHOOTER spectrum (grey) and the best-fit model (red) of \sd. The projected rotational velocity is set to \vsini\ =  20 \kms. 
Line positions at zero magnetic field are labelled.}
\label{fig:Hdelta}
\end{figure*}

We performed a global $\chi^2$ fit to the X-SHOOTER spectrum of \sd\ to determine its atmospheric parameters, including $B$ and $\psi$. 
Some of the strongest hydrogen and helium lines that could be used for the fit are shown in Fig.~\ref{fig:atmlines}.
The $\chi ^2$ fit prefers a projected rotational velocity of $v_\mathrm{rot}\sin i=34$\,\kms. 
However, \vsini\ is not well constrained because broadening of the displaced Zeeman components may also be caused by a more complex magnetic field geometry. 
Possibly for the same reason, the central components of several helium lines are sharper in the observation than predicted by our model. 
The strength of the central components of Zeeman triplets may also be somewhat increased by magneto-optical effects \citep{Martin1981}, which are not included in our models. 
Most observed metal lines appear sharp given the limited resolution of the X-SHOOTER spectra. 
The final atmospheric parameters as derived from the X-SHOOTER spectra are listed in the left part of Table \ref{tab:results}.
All uncertainties are estimated because they are dominated by systematic effects rather than by noise.

As demonstrated in Fig.~\ref{fig:Hdelta}, the X-SHOOTER spectrum of \sd\ shows strong \ion{N}{iii-iv}, \ion{Si}{iv}, and \ion{S}{iv} lines.
The determination of abundances was complicated not only by the magnetic field but also because most transitions originate from high-lying energy levels that are hard to model, even in non-magnetic sdO stars.
This meant that some predicted lines could not be used for the abundance determination, for example the \ion{C}{iii} triplet at 4070\,\AA\ and \ion{N}{iii}\,4379\,\AA. 
Metal abundances were therefore estimated by eye, by comparing the observed spectra with models with different abundances and keeping the atmospheric parameters fixed to the best-fit values. 
The derived abundances are listed on the right side of Table \ref{tab:results}. %
\sd\ is enriched in nitrogen at close to 20 times the solar abundance (by number). 
Carbon and oxygen seem to be at least somewhat subsolar because no lines from these elements are clearly detected -- in particular, the \ion{C}{iii} 4159\,\AA, \ion{C}{iv} 5805\,\AA, and \ion{O}{iii} 3760\,\AA\ multiplets.
The photosphere therefore likely consists of material processed by hydrogen fusion in the CNO cycle. 
Silicon and sulphur are enriched at about two and six times their solar value, respectively.

The X-SHOOTER spectra give us a precise radial velocity for \sd. 
To test whether \sd\ is radial velocity variable or not, follow-up spectroscopic observations are necessary because %
the IDS spectra are of insufficient quality.

By combining the spectroscopic surface gravity, the \textit{Gaia}  parallax, and the spectral energy distribution (SED), the mass, radius, and luminosity of \sd\ can be derived. This analysis is described in Appendix \ref{sect:sed}. 
While the stellar radius, $0.184\pm0.011$\,$R_\odot$, and luminosity, $123^{+19}_{-16}$\,$L_\odot$, are typical for He-sdO stars, the derived mass for \sd, $0.93^{+0.44}_{-0.30}$\,$M_\odot$, is rather high. 
It should be noted that the uncertainty of the mass is considerable due the limited accuracy of the spectroscopic surface gravity. 
These parameters place \sd\ close to the zero-age helium main sequence at $\approx$0.8\,\msun\ \citep{Paczynski1971}, which means it consists of a helium core with next to no hydrogen envelope left.

\section{Evolutionary status}
\label{sect:evo}

In general, the presence of a strong magnetic field can be considered the smoking gun for stellar mergers \citep{schneider2019}.
Given that \sd\ is an apparently single magnetic helium-rich sdO star and likely more massive than usual for that class of star \citep{schindewolf2018,dorsch2019,dorsch2020,dorsch2021}, the merging of two He-WD stars can be considered a natural scenario for its formation. 
In this scenario an object with a mass as high as 0.93\,\msun\ would be unlikely, but the scenario is predicted to produce significant numbers of He-sdO stars with masses of up to about 0.8\,\msun\ \citep{Han2003}, which would be well within the uncertainty of our mass for \sd.
A related formation channel for He-sdO stars is the merger of an He-WD with a post-sdB WD, which would produce a hybrid WD with a CO core and a thick helium envelope, a so-called HeCO-WD (\citealt{Justham+2011}, following pioneering models from \citealt{Iben1990}). This channel is a natural consequence of one of the common-envelope formation channels for sdB stars described by \citet{han02} and seems likely to contribute substantially to the population of single He-sdO stars. Moreover, the inferred effective temperature and surface gravity of \sd\ are close to those of the densest region of the population predictions in \citet{Justham+2011}.

Based on the three-dimensional simulations of CO WD mergers of \cite{yoon2007} and \cite{loren2009}, \citet{garcia2012} showed that strong magnetic fields can be generated in the hot, convective, differentially rotating corona that surrounds the primary during and shortly after the merging. These magnetic fields are predicted to be frozen to the outer layers of the final merger product. Magneto-dipole radiation rapidly spins down the newly formed magnetic WD if magnetic and rotation axes are not aligned.
These predictions appear to be consistent with the recent results of \cite{Bagnulo2021}, who found that young WDs with fields exceeding 1\,MG are more massive than canonical WDs.  

\begin{figure}
\centering
\includegraphics[width=0.99\columnwidth]{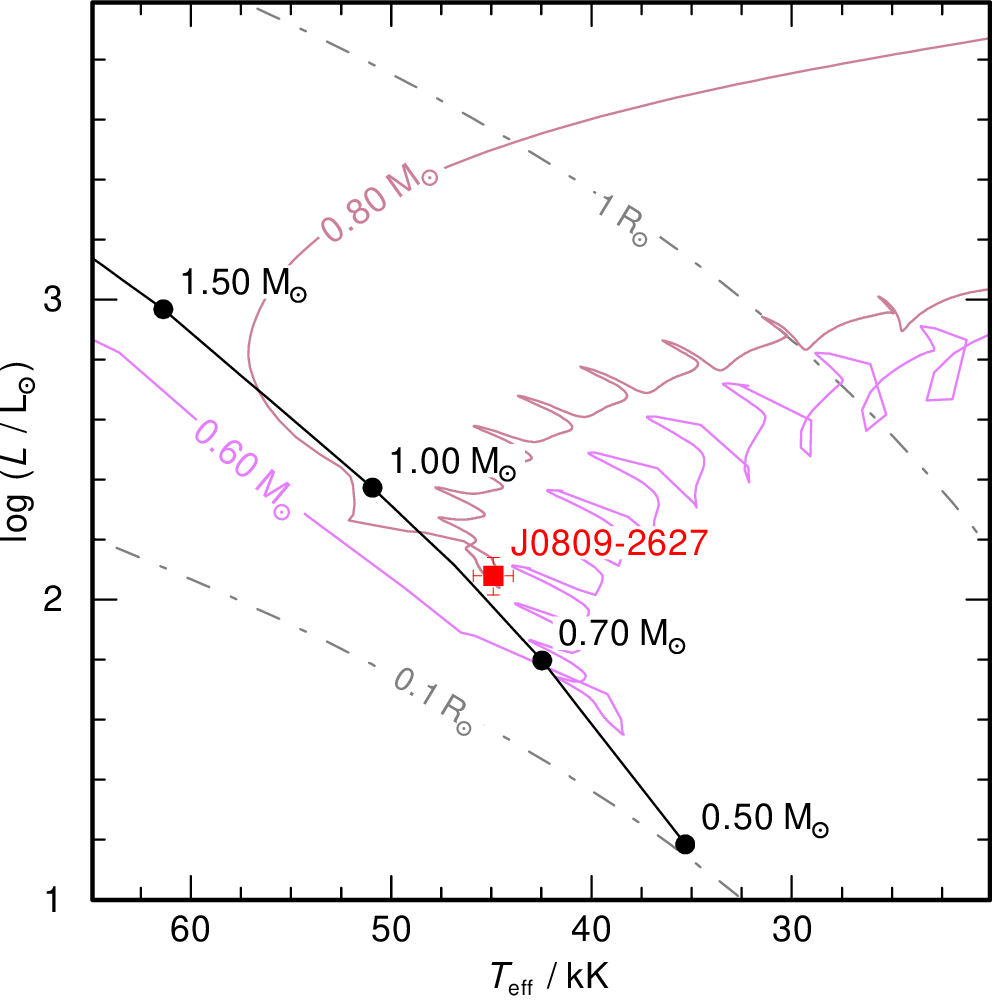}
\caption{Position of \sd\ (red) in the Hertzsprung-Russell diagram compared to several evolutionary tracks. 
Pink lines show two double He-WD slow-merger tracks from \cite{zhang12a} for metallicity $Z=0.02$.
The final masses for both tracks are labelled.
Black lines show the zero-age helium main sequence from \cite{Paczynski1971}, where black filled circles indicate masses of 0.5, 0.7, 1.0, and 1.5 \msun.
The dashed-dotted grey lines indicate radii of 0.1 and 1\,$R_{\odot}$.
}
\label{fig:hrd}
\end{figure}

The CNO-processed chemical composition and the relatively slow projected rotation of \sd\ provide additional important constraints on merger models.  
Evolutionary calculations for mergers of He-WDs with He-WDs  were performed by \cite{zhang12a}, \cite{Hall2016}, and \cite{Schwab2018}.
\cite{zhang12a} predict nitrogen-rich surfaces for slow double He-WD mergers, similar to what we find for \sd. 
Such systems retain the initial composition of the secondary He-WD, which includes a CNO-cycle signature, as observed for \sd. 
Because our upper limit for the carbon abundance of \sd\ is only slightly subsolar, the composite merger model of \cite{zhang12a} cannot be excluded for remnant masses below about 0.7\,\msun.
This model assumes that more than half of the secondary mass is rapidly accreted and forms a hot corona, after which the remaining mass is accreted slowly.
As shown in Fig.~\ref{fig:hrd}, the 0.8\,\msun\ slow merger model of \cite{zhang12a} matches the effective temperature and luminosity of \sd\ well at the onset of core helium burning.

For hybrid WD mergers, it seems plausible that surface abundances consistent with \sd\ are also possible as long as the He-WD is disrupted rather than the hybrid HeCO-WD (in other words, as long as the He-WD is the less massive component of the merging binary). Of the hybrid merger models in \citet{Justham+2011}, the luminosity and surface temperature of \sd\ are matched by mergers in which a hybrid WD of only 0.35\,\msun\ accretes 0.15--0.3\,\msun\ from a merger with a He-WD.  Those models are also consistent with the surface gravity of \sd. However, only the more massive of those merger models are within the mass range we infer for \sd.

The best-fit hydrogen abundance for \sd\ is higher than that observed for most (extreme) He-sdO stars \citep{stroeer07,schindewolf2018}. Hydrogen was not included in the merger models of \citet{Justham+2011} or \citet{zhang12a}, while both hydrogen and helium were included in the models of \cite{Schwab2018}.
The \cite{Schwab2018} models predict that the surface hydrogen abundance depends on the rotational velocity after the merging -- models that include rotation result in very hydrogen-poor surfaces, which is not what we find for \sd.
Specifically, \cite{Schwab2018} predict rotational velocities of about 30\,\kms\ once a 0.5\,\msun\ merger remnant reaches the core-helium-burning phase, which would be consistent with the elemental abundance pattern of \sd. 
However, a more massive 0.7\,\msun\ merger remnant is predicted to be a fast rotator, with $v_\mathrm{rot}$ $\approx$ 100\,\kms.
This would require a low inclination for \sd\ given its low projected rotational velocity ($<$\,40\,\kms).
Alternatively, a misalignment of rotation and magnetic axes might provide further loss of angular momentum through magneto-dipole radiation \citep{garcia2012}, which was not included in the calculations of \cite{Schwab2018}. 
In the 0.7\,\msun\ model of \cite{Schwab2018}, the surface is predicted to be enriched in carbon, which, however, is not observed for \sd.

\sd\ will evolve through a helium-shell-burning phase and eventually become a WD. 
If the magnetic flux is conserved until \sd\ reaches the WD stage ($B\sim R^{-2}$), its surface field strength will reach about 120\,MG at a radius of 0.01\,\rsun.

\section{Summary and conclusions}

We performed a detailed spectral analysis of the magnetic He-sdO \sd\ and used a simple homogeneous model for the magnetic field to derive a mean magnetic field strength of $353\pm10$\,kG at an inclination of $\psi = 64 \pm 25\,\degr$. 
Apart from its strong magnetic field, \sd\ has a super-solar helium abundance at \logy = $+0.28\pm0.10$, which is low compared to typical non-magnetic He-sdO stars. Its effective temperature, \teff\ = $44900\pm 1000$\,K, and surface gravity, \logg\ = $5.93\pm 0.15$, are not unusual. 
The He-sdO is nitrogen-rich and seems to be carbon- and oxygen-poor, although only upper limits could be derived for C and O. 
We combined astrometry, photometry, and spectroscopy to derive stellar parameters of $R$~=~$0.184\pm0.011$\,$R_\odot$, $L$~=~$123^{+19}_{-16}$\,$L_\odot$, and $M$~=~$0.93^{+0.44}_{-0.30}$\,$M_\odot$. %
Placing it in the Hertzsprung-Russell diagram, \sd\ is located on the helium main sequence, which implies that it has completely lost any hydrogen envelope.

The \textit{TESS} light curve of \sd\ does not show significant variability (see Appendix \ref{sect:appendix_obs_tess}).
Assuming a radius of $0.184$\,$R_\odot$ and a rotational velocity of less than 40\,\kms, a magnetic spot would be expected to result in a modulation at a period of more than 5.6\,h. The observed lack of photometric variability may be caused by a low inclination of the rotational axis, although a low inclination of the magnetic field axis seems to be excluded by spectroscopy. However, many magnetic main sequence stars are oblique rotators, which means that their magnetic axis is not aligned with their rotational axis \citep{landstreet1987}. 
Photometric observations that are more sensitive than the $\approx$0.42\,ppt detection limit achieved by the current TESS data would help to further constrain the presence of magnetic spots.

In summary, the detection of the strong magnetic field of \sd, its unusually high mass, its chemical composition, its slow projected rotation, and the lack of evidence for a close stellar companion provide overwhelming evidence for \sd\ being the result of a double-degenerate merger that produced a stably helium-burning star.
Nonetheless, radial velocity monitoring still has to be done to exclude a close binary nature\footnote{He-sdOs with masses as high as that of \sd\ or more could also form from intermediate mass stars that had their envelope stripped by a compact companion \citep{Goetberg2018}. However, the stripping process is unlikely to generate a strong magnetic field, so the magnetic field would need to be primordial or generated during the red giant evolution of the progenitor. In addition, the stripped stars of \cite{Goetberg2018} are predicted to be much more luminous than \sd.}. As noted by \cite{Schwab2018}, magneto-hydrodynamic simulations would be required for suitable double He-WD mergers or He-WD+HeCO-WD mergers, similar to those performed for CO-WD mergers by \cite{Ji2013} and \cite{Zhu2015}. 

The double He-WD and He-WD+HeCO-WD merger scenarios are also both viable for the more common non-magnetic He-sdO stars as well. Therefore, the question remains as to why the vast majority of He-sdO stars are apparently non-magnetic while \sd\ has a strong magnetic field. 
One possibility would be that the surface magnetic field observed for \sd\ is not stable and weakens quickly after a merging event. 

While our simple homogeneous model for the magnetic field generally results in good fits for most lines in the X-SHOOTER spectrum of \sd, the magnetic field geometry might be more complicated in reality.
Spectropolarimetric follow-up observations would provide more information about the magnetic field geometry. 
The current X-SHOOTER spectrum excludes projected rotation velocities of more than about 40\,\kms, which is still consistent with rotation velocities predicted for some double He-WD merger remnants \citep{Schwab2018}.
Higher-resolution monitoring would further constrain this value.
Ultraviolet observations would allow us to improve on the determination of abundances of carbon and oxygen, for which only upper limits are available at present, and would also provide abundance measurements for many more metals, including the iron group. A complete abundance pattern is crucial for establishing the composition of the accreted material.

\begin{acknowledgements}

We thank the referee, John Landstreet, for his very useful comments. 
We thank Ylva G\"{o}tberg for discussions on stripped massive stars and Max Pritzkuleit for his support with the HOTFUSS project.
UH and MD acknowledge funding by the Deutsche For\-schungs\-gemeinschaft (DFG) through grants IR190/1-1, HE1356/70-1 and HE1356/71-1.
IP acknowledges support from the UK's Science and Technology Facilities Council (STFC), grant ST/T000406/1.
AGI acknowledges support from the Netherlands Organisation for Scientific Research (NWO). SJ acknowledges funding from the Netherlands Organisation for Scientific Research (NWO), as part of the Vidi research program BinWaves (project number 639.042.728, PI: de Mink).
This research has made use of NASA's Astrophysics Data System. Based on observations made with ESO Telescopes at the La Silla Paranal Observatory under programme ID 105.206H.001 and observations with the Isaac Newton Telescope (under programme ID ING.NL.19B.005) operated on the island of La Palma by the Isaac Newton Group of Telescopes in the Spanish Observatorio del Roque de los Muchachos of the Instituto de Astrofísica de Canarias. 
This paper includes data collected by the TESS mission. Funding for the TESS mission is provided by the NASA's Science Mission Directorate.
This work has made use of data from the European Space Agency (ESA) mission {\it Gaia} (\url{https://www.cosmos.esa.int/gaia}), processed by the {\it Gaia} Data Processing and Analysis Consortium (DPAC, \url{https://www.cosmos.esa.int/web/gaia/dpac/consortium}). Funding for the DPAC has been provided by national institutions, in particular the institutions participating in the {\it Gaia} Multilateral Agreement.

\end{acknowledgements}

\bibliographystyle{aa}
\bibliography{mHe-sdO.bib}

\begin{appendix}
\label{appendix}

\section{Observations}\label{sect:appendix_obs}

\subsection{Spectroscopy}\label{sect:appendix_obs_spec}

Details of the spectroscopic observations are listed in Table \ref{tab:observations}. 

\begin{table}
\centering
\caption{Dates and exposure times for the spectroscopic observations of \sd.}
\label{tab:observations}
\resizebox{\columnwidth}{!}{
\begin{tabular}{lcc}
\toprule
\toprule
Spectrograph & Mid exposure BJD & Exposure time (s) \\
\midrule
INT/IDS      &   2458837.6695        &  750  \\
INT/IDS      &   2458837.6798        &  750  \\
VLT/X-SHOOTER &   2459314.5691        &  1800 \\
\bottomrule
\end{tabular}
}
\end{table}

\subsection{\textit{TESS}}\label{sect:appendix_obs_tess}

\begin{figure}
\centering
\includegraphics[width=\columnwidth]{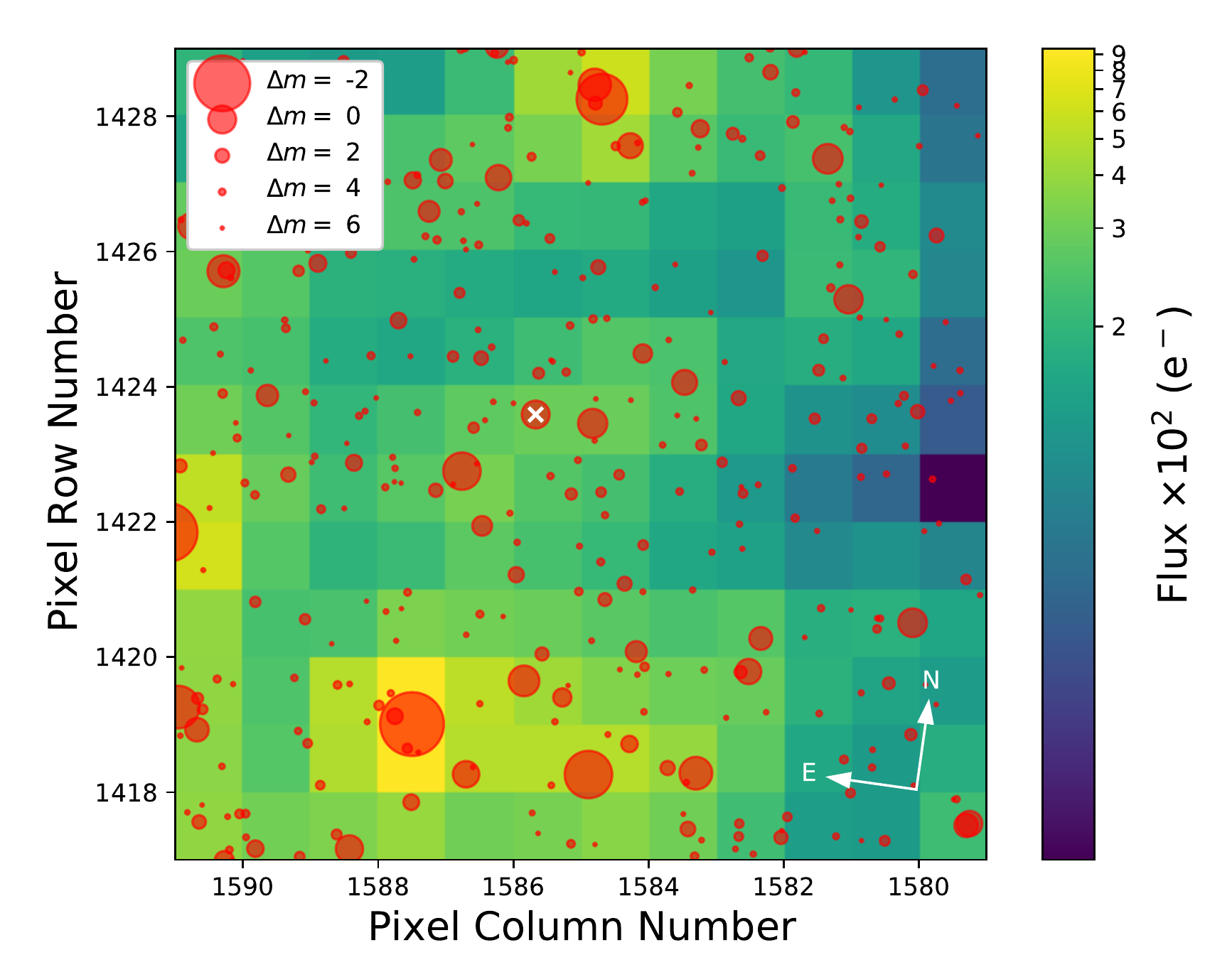}
\caption{Image from \textit{TESS} sector 7 created with {\tt tpfplotter} \citep{tpfplotter} showing the location of \sd\ (white cross) and other stars with a magnitude difference of six or less (red circles).}
\label{fig:TPF}
\end{figure}

\begin{figure}
\centering
\includegraphics[width=\columnwidth]{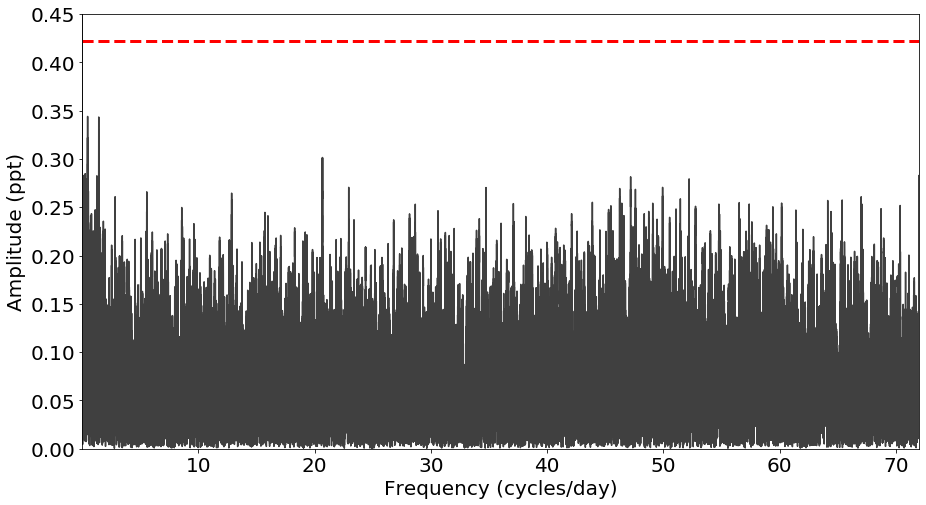}
\caption{Fourier transform of the \textit{TESS} data. The amplitude is shown in ppt. The adopted detection limit of 0.42~ppt is shown as a dashed red line, and no peaks are detected above this limit.}
\label{fig:TESS}
\end{figure}

\sd\ was in the \textit{TESS} field during sectors 7, 8, and 34. The star was not targeted for short-cadence observations, but the full-frame images are available. They have an integration time of 30~min for sectors 7 and 8, and 10~min for sector 34. The total time span of the light curve is 760~days. We performed the photometry of all available images using {\tt eleanor} \citep{eleanor}. As the field is relatively crowded (see Fig.~\ref{fig:TPF}), we used the small aperture option, which limits aperture size to 8 pixels\footnote{The \textit{TESS} pixel size is 21''.}. We then carried out the Fourier transform of all the data up to the Nyquist frequency, as shown in Fig.~\ref{fig:TESS}. In order to determine a detection limit, we randomly assigned one of the
measured fluxes to each measured time and recalculated the Fourier transform a thousand times, each time recording the maximum amplitude. We selected the maximum overall value as the detection limit; as it happens once every 1000 runs, this corresponds to a false-alarm probability of $1/1000$. No peaks are detected above this threshold; therefore, we find no periodic variability for periods between $\approx 20$~min and $\approx 380$~days down to a detection limit of $\approx 0.42$~ppt. It is important to emphasise, however, that the large pixel size can result in contamination from nearby constant stars, which can hide any variability. Follow-up with better image resolution is encouraged.

\section{Linear Zeeman effect}
\label{appendix:zeeman}

We assume LS coupling for all metal line transitions given that the heaviest observed element is sulphur. An LS coupling state is described by its orbital angular momentum, $L$, its spin angular momentum, $S$, and its total angular momentum, $J$.
For each included line, the upper and lower LS states were split into $2J+1$ components with magnetic quantum numbers from $m=-J$ to $m=+J$. Only transitions with $\Delta m = 0$ or $\pm$1 were considered. 

For a transition between lower and upper states with magnetic quantum numbers $m_\mathrm{l}$ and $m_\mathrm{u}$, the wavelength shift with respect to the rest wavelength at zero magnetic field, $\lambda _0$, is given as
\begin{equation}
\Delta \lambda = \frac{eB\lambda_0^2}{4\pi m_e c^2} \cdot (m_\mathrm{l}g_\mathrm{l}-m_\mathrm{u}g_\mathrm{u})
,\end{equation}
where $e$ is the elementary charge, $B$ is the magnetic field strength, $m_e$ is the electron mass, $c$ is the speed of light, and $g_\mathrm{l}$ and $g_\mathrm{u}$ are the Landé $g$-factors for the lower and upper level.

In the case of LS coupling, the Landé $g$-factor is given as
\begin{equation}
g = 1+\frac{J(J+1)+S(S+1)-L(L+1)}{2J(J+1)}.
\end{equation}
Transitions between levels of hydrogen and singly ionised helium represent special cases in this model for the magnetic field because these levels are sufficiently described by a main quantum number, $n$.
For these levels, as well as neutral helium, the Landé $g$-factors were set to unity.

\begin{table}
\centering
\caption{Relative intensities for the $\pi$ and $\sigma$ components following \cite{Hoenl1925}. The quantum numbers $J$ and $m$ refer to the initial level, and $\Delta J = J_\mathrm{final} - J_\mathrm{initial}$.
We note the difference from Eq.~(5) of \cite{Kawka2011}, which contains a sign error.
}
\label{tab:rint}
\begin{tabular}{ccr}
\toprule
\toprule
$\Delta J$ & $\pi$ & $\pm \sigma$  \\
\midrule
$0$  & $m^2$ & $\frac{1}{4}(J \mp m)(J\pm m +1)$  \\
$+1$ & $(J+1)^2-m^2$ & $\frac{1}{4}(J \pm m +1)(J\pm m +2)$  \\
$-1$ & $J^2-m^2$  & $\frac{1}{4}(J \mp m)(J\mp m -1)$ \\
\bottomrule
\end{tabular}
\end{table}

Analytic expressions for the relative intensities of the components of any linear Zeeman multiplet were first presented by \cite{Hoenl1925}.
They are summarised in Table~\ref{tab:rint}.
Unlike the Landé $g$-factors, these relative intensities depend only on $J$ and $m$, and are therefore valid in any coupling scheme.

The relative intensity ($I$) of the central ($\pi$) and displaced ($\sigma$) components depends on the angle ($\psi$) between the magnetic field axis and the line of sight.
The angular dependence given by \cite{Condon1935} is
for the $\pi$ (central) component
\begin{equation}
\frac{I(\psi)}{I(90\degr)}   = \sin ^2 \psi
\end{equation}
and for the $\sigma$ components
\begin{equation}
\frac{I(\psi)}{I(90\degr)} = 1+\cos ^2 \psi.
\end{equation}

\section{Mass, radius, luminosity, and stellar population characteristics}
\label{sect:sed}

\begin{figure}
\centering
\includegraphics[width=0.49\textwidth]{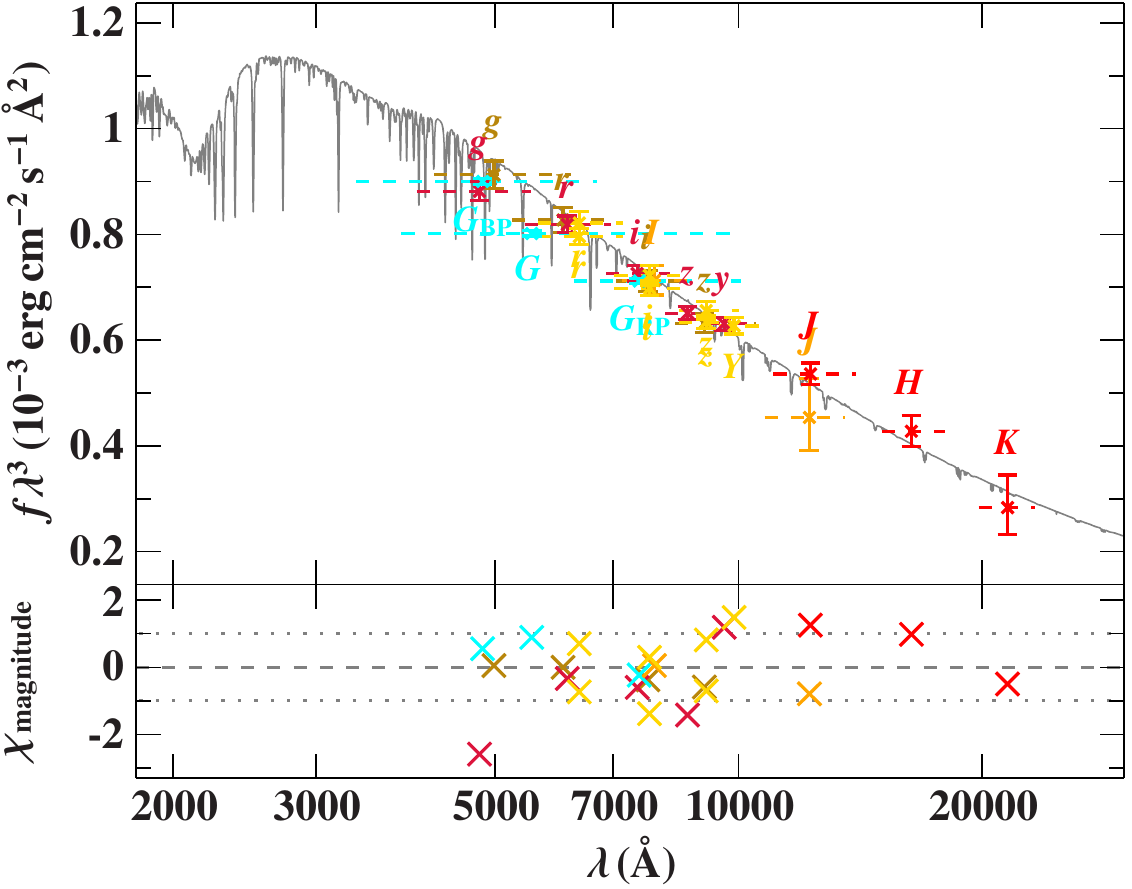}
\caption{SED fit for \sd. The grey line shows the final model of \sd,\ and filter-averaged flux measurements are indicated by dashed horizontal lines. Photometric surveys are identified by colour codes: 
Pan-STARRS \citep[dark red;][]{Magnier2020},
SkyMapper \cite[dark gold;][]{Onken2019},
\textit{Gaia} \citep[cyan;][]{Riello2020}, 
DECam \citep[gold;][]{Schlafly2018,Drlica-Wagner2021},
DENIS \cite[orange;][]{vizier:B/denis}, and 
2MASS \cite[red;][]{Cutri2003_2MASS}.
}
\label{fig:SED}
\end{figure}

\begin{table}
\centering
\caption{SED fit and stellar parameters. The stellar parameters result from the SED fit with a prescribed \teff\ = $44900 \pm 1000$\,K. 
The mode and the highest density interval of each quantity are given for 1$\sigma$ probability \citep[see][]{bailer18}.}
\label{tab:SED}
\setstretch{1.3}
\begin{tabular}{lr}
\toprule
\toprule
{} & SED fit  \\
\midrule
\teff\ (K) & $47000^{+11000}_{-3000}$ \\
\midrule
$\log \Theta\,\mathrm{(rad)}$ & $-11.246^{+0.005}_{-0.004}$  \\
$E_{44-55}$ (mag) & $0.073^{+0.004}_{-0.004}$ \\
Radius $R = \Theta/(2\varpi)$  & $0.184^{+0.011}_{-0.010}$\,$R_\odot$ \\
Mass $M = g R^2/G$  & $0.93^{+0.44}_{-0.30}$\,$M_\odot$ \\
Luminosity $L/L_\odot = (R/R_\odot)^2(T_\mathrm{eff}/T_{\mathrm{eff},\odot})^4$  & $123^{+19}_{-16}$ \\
\bottomrule
\end{tabular}
\end{table}

We constructed the SED of \sd\ by collecting photometric measurements from several surveys. 
Two SED fits were then performed using the same grid as for the spectroscopic analysis, with \teff\ either as a free parameter or fixed to the value derived by the spectroscopic fit \citep[for details, see][]{Heber2018,irrgang2021}. 
The latter approach is useful because \teff\ $\gtrsim$ 35\,000\,K are not well constrained from optical and infrared magnitudes alone. 
The surface gravity and helium abundance were always fixed to spectroscopic values, and the angular diameter $\Theta$ was always a free parameter. 
Interstellar extinction was accounted for using the reddening law of \cite{Fitzpatrick2019} with the colour excess $E_{44-55}$ as a free parameter. An extinction parameter of $R (55) = 3.02$ was assumed. 

The angular diameter derived from the SED (with fixed \teff) was combined with the parallax measurement provided by \textit{Gaia} Early Data Release 3 \citep[EDR3;][]{Gaia2016,Gaia2020} and the spectroscopic surface gravity to estimate the stellar parameters mass, radius, and luminosity for \sd. 
Here, the parallax measurement, $\varpi=0.68\pm 0.04$\,mas, was corrected for its zero point offset following \cite{Lindegren2021}, and the corresponding uncertainty was inflated using the function of \cite{El-Badry2021}.
The SED fit and stellar parameters are summarised in Table \ref{tab:SED}. 
Figure \ref{fig:SED} shows the SED fit with \teff\ fixed to the spectroscopic value.

Because \sd\ is close to the Galactic disk, we expect it to belong to the thin disk stellar population I. This can be confirmed by calculating its current Galactic space velocity vector from the radial velocity (assumed to be non-variable), parallax, and proper motions provided by \textit{Gaia} EDR3 \citep{Gaia2020}. The resulting vector ($U$, $V$, $W$) = ($-17.3$ $\pm$ 2.0, 230.1 $\pm$ 2.7, 7.1 $\pm$ 0.8) \kms, where $U$ is measured towards the Galactic centre, $V$ in the direction of Galactic rotation, and $W$ perpendicular to the disc, is consistent with a membership of the thin disk population.

\section{Spectral comparisons}

Here we show spectral comparisons for the X-SHOOTER spectra of \sd, specifically the UVB and VIS spectral arms up to 7400\,\AA.
The strongest lines are labelled at their position for zero magnetic field. Spectral regions that are not well reproduced by our models were excluded from the fit. 
This includes several metal lines, as well as some neutral helium lines.
Especially narrow \ion{He}{i} lines from the triplet ($S$=1) system, such as \ion{He}{i} 5876\,\AA, and lines with strong forbidden components, such as \ion{He}{i} 4472, 4922\,\AA, are not well modelled.
This is to be expected given the simple nature of our model for the magnetic field, as well as the lack of helium broadening tables in the presence of a magnetic field.
The \ion{C}{iii} lines at 4070\,\AA\ are poorly modelled due to limitations of the \textsc{Tlusty} model ion. 
Telluric lines were removed using the grid of transmission spectra provided by \cite{Moehler2014}.

\begin{figure*}
\centering
\includegraphics[width=24.2cm,angle=90,page=2]{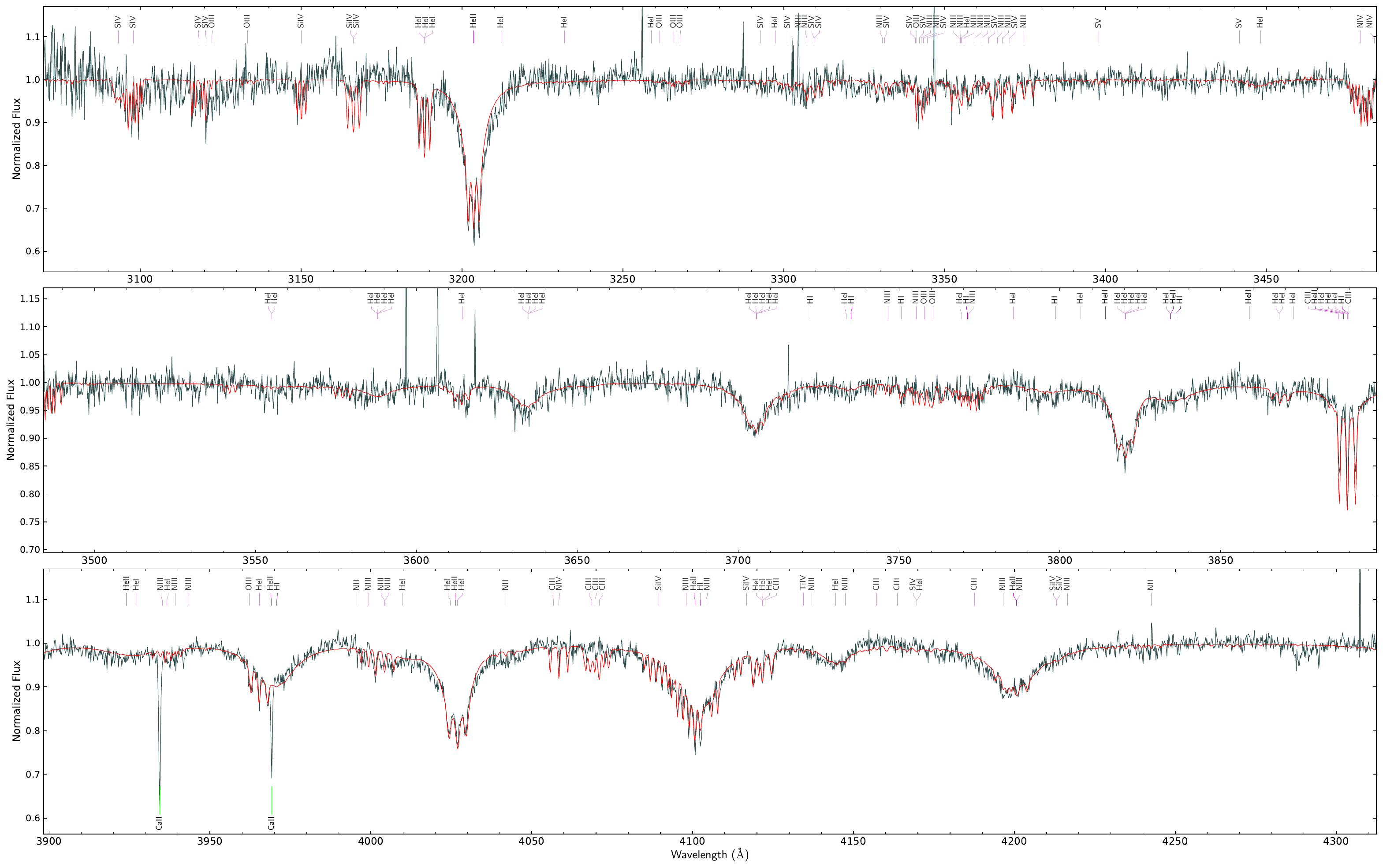}
\caption{X-SHOOTER spectrum of \sd\ (grey) and the best-fit model (red).}
\label{appendix:xshooter}
\end{figure*}
\captionsetup[ContinuedFloat]{labelformat=continued}
\begin{figure*}
\ContinuedFloat
\centering
\includegraphics[width=24.2cm,angle=90,page=3]{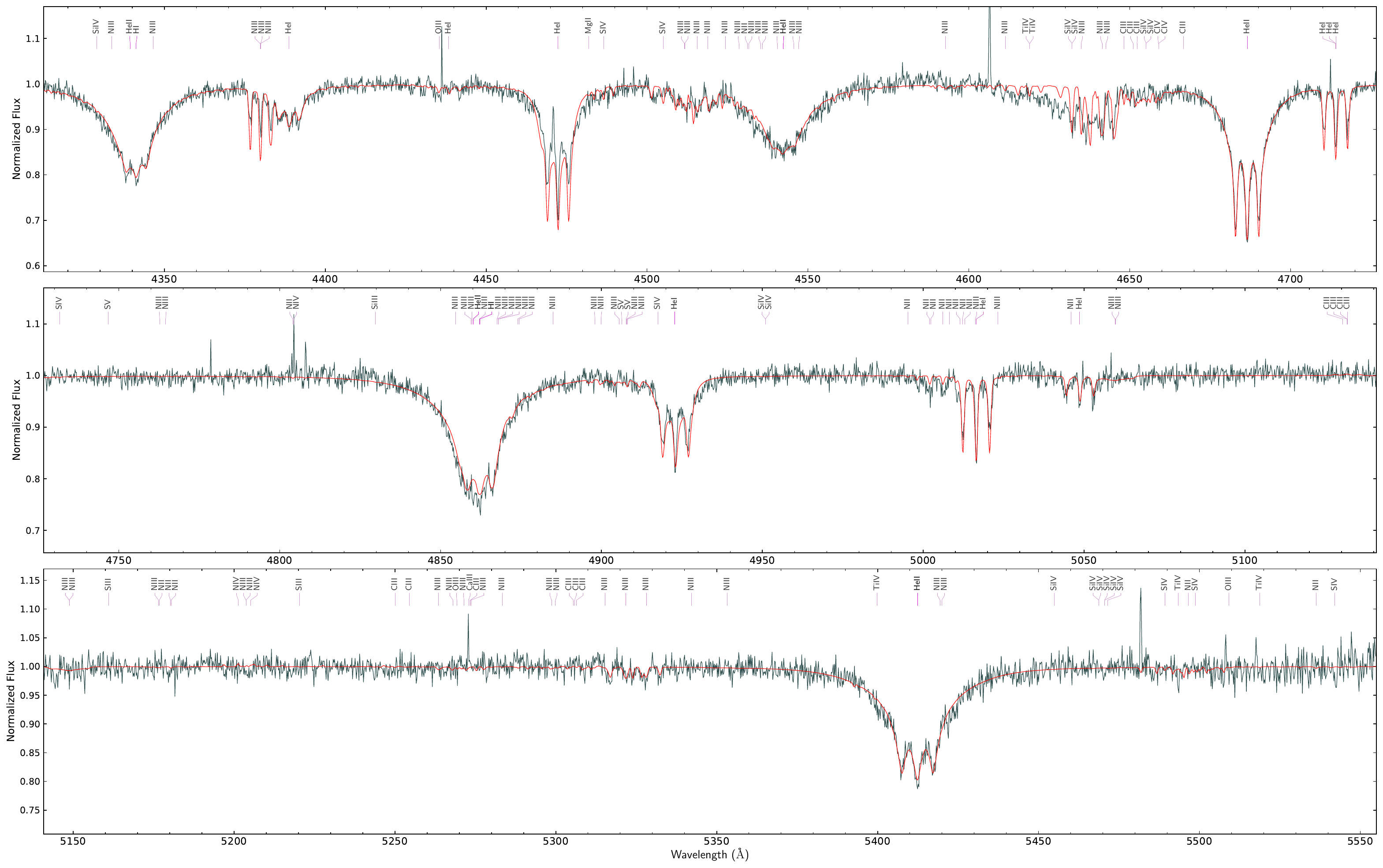}
\caption{X-SHOOTER spectrum of \sd\ (grey) and the best-fit model (red).}
\end{figure*}
\begin{figure*}
\ContinuedFloat
\centering
\includegraphics[width=24.2cm,angle=90,page=2]{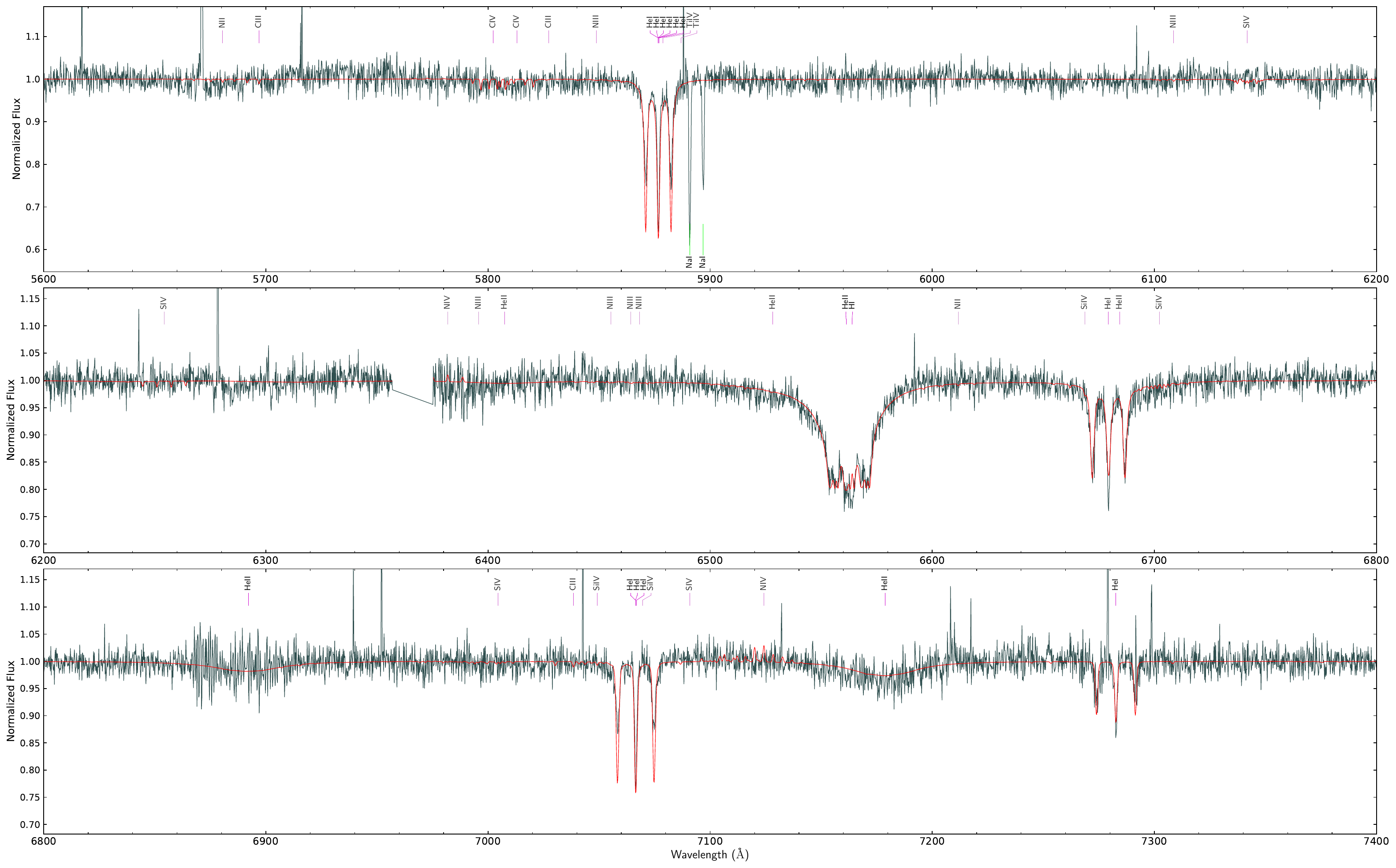}
\caption{X-SHOOTER spectrum of \sd\ (grey) and the best-fit model (red).}
\end{figure*}

\end{appendix}

\end{document}